# Initial demonstration of AlGaAs-GaAsP-β-Ga₂O₃ n-p-n double heterojunctions

Jie Zhou[1], Ashok Dheenan[2], Jiarui Gong[1], Carolina Adamo[3], Patrick Marshall[3], Moheb Sheikhi[1], Tsung-Han Tsai[1], Nathan Wriedt[2], Clincy Cheung[3], Shuoyang Qiu[1], Tien Khee Ng[4], Qiaoqiang Gan[4], Gambin Vincent[3], Boon S. Ooi[4], Siddharth Rajan[2], and Zhenqiang Ma[1]*

*[1]Department of Electrical and Computer Engineering, University of Wisconsin-Madison, Madison, Wisconsin, 53706, USA*

*[2]Department of Electrical and Computer Engineering, The Ohio State University, Columbus, OH 43210, USA*

*[3]Northrop Grumman, Redondo Beach, CA 90278, USA*

*[4]Department of Electrical and Computer Engineering, King Abdullah University of Science and Technology, Thuwal 23955-6900, Saudi Arabia*

E-mail: mazq@engr.wisc.edu, rajan.21@osu.edu, boon.ooi@kaust.edu.sa

β-Ga₂O₃, an ultrawide-bandgap semiconductor, has great potential for future power and RF electronics applications but faces challenges in bipolar device applications due to the lack of p-type dopants. In this work, we demonstrate monocrystalline AlGaAs/GaAsP/β-Ga₂O₃ n-p-n double-heterojunctions, synthesized using semiconductor grafting technology. By transfer printing an n-AlGaAs/p-GaAsP nanomembrane to the n-β-Ga₂O₃ epitaxial substrate, we simultaneously achieved AlGaAs/GaAsP epitaxial n-p junction diode with an ideality factor of 1.29 and a rectification ratio of $2.57 \times 10^3$ at +/-2 V, and grafted GaAsP/β-Ga₂O₃ p-n junction diode exhibiting an ideality factor of 1.36 and a rectification ratio of $4.85 \times 10^2$ at +/- 2 V.

Beta-phase gallium oxide (β-Ga$_2$O$_3$) is gaining attention in the semiconductor sector due to its superior electronic properties, such as an ultrawide bandgap of 4.9 eV, a high breakdown field of 10.3 MeV, and a high electron saturation velocity of $1.1 \times 10^7$ cm/s.[1]–[3] These attributes, coupled with the recent advancement in the epitaxy growth technology and accessibility to large diameter native substrates, make β-Ga$_2$O$_3$ a promising candidate for next-generation high-speed, high-power electronics as well as solar-blind optoelectronics.

Despite the fast developments of unipolar β-Ga$_2$O$_3$ devices, such as Schottky diodes[4]–[6], field-effect transistors[7]–[9], and photodetectors[10]–[12], a primary impediment for broader β-Ga$_2$O$_3$ device applications is the limited bipolar performance as a result of inefficient acceptor dopants for β-Ga$_2$O$_3$.[13]–[15] In general, the candidate p-type dopants introduce deep acceptor levels in β-Ga$_2$O$_3$ bandgap, rendering them ineffective in generating free holes for conduction due to their elevated ionization energies. Various strategies have been proposed to address or circumvent this p-type inefficiency, as elaborated in multiple studies. For example, surface activated/direct wafer bonding has been employed to hetero-integrate β-Ga$_2$O$_3$ with other materials.[16]–[20] Deposition of p-type polycrystalline oxides such as NiO, Cu$_2$O on n-type β-Ga$_2$O$_3$ provides another perspective to create high-performance heterojunction devices.[21]–[23] In the meantime, utilizing micro-transfer printing technique, Zheng *et al* demonstrated Si/β-Ga$_2$O$_3$ heterojunction photodetectors, which exhibited improved photoresponsivity and quantum efficiency.[24] All these prior works have advanced the β-Ga$_2$O$_3$-based bipolar devices; however, these methods still face challenges such as imperfect surfaces, intermixing of atoms, polycrystallinity, and excess interfacial traps. which may pose new challenges in further developing β-Ga$_2$O$_3$-based bipolar devices in the future. Exploring new approaches to developing β-Ga$_2$O$_3$-based bipolar devices that can overcome some, if not all, of the challenges, is needed.

Recently, a novel method utilizing semiconductor grafting technology has been introduced to create high-quality monocrystalline Si/β-Ga$_2$O$_3$ heterojunction p-n diodes.[25] Semiconductor grafting allows for the creation of abrupt heterojunctions with epitaxy-like interfaces between semiconductors, irrespective of lattice structures or constants.[26]–[29] Unlike the conventional wafer bonding/fusion or direct deposition/transfer printing, semiconductor grafting incorporates an ultrathin oxide (UO) interlayer between two single crystalline semiconductors. This UO layer, typically sub-nanometer thick, can be introduced extrinsically through atomic layer deposition (ALD), or intrinsically via controlled interfacial oxidation. Both approaches create a well-passivated interface, suppressing the density of states and ensuring efficient charge transport across the junction through quantum tunneling.

In this work, we present the integration of two dissimilar semiconductor material systems, GaAs-based III-Vs (consisting of GaAs/AlGaAs/GaAsP triple layers) and β-$Ga_2O_3$, using semiconductor grafting technology. As a result, a double-heterojunction structure, specifically, the n-p-n AlGaAs-GaAsP-β-$Ga_2O_3$, was synthesized. The first heterojunction (HJ1) of n-p AlGaAs/GaAsP was formed epitaxially through conventional heteroepitaxy and was subsequently grafted to a n-type β-$Ga_2O_3$ epitaxial substrate. This in turn leads to a second heterojunction (HJ2) of p-n GaAsP/β-$Ga_2O_3$, interfaced by a sub-nanometer thick $Al_2O_3$ layer. Both heterojunctions were fabricated into diodes, and current-voltage (I-V) characterizations reveal high-performance rectifying behaviors in terms of ideality factors and rectification ratios. The direct comparison between the two junctions, one formed by epitaxy and the other via grafting, suggests that the lattice-mismatched heterojunction created by semiconductor grafting can rival the quality of lattice-matched epitaxy. The advancement could facilitate the construction of more complicated device architectures, such as heterojunction bipolar transistors (HBTs), and could eventually harness the full electronic potential of β-$Ga_2O_3$.

The schematic of the synthesized n-p-n double-heterojunction structure is depicted in Fig. 1(a). The first epitaxial heterojunction (HJ1) consists of a 180 nm n-type $Al_{0.3}Ga_{0.7}As$ and 80 nm p-type GaAsP. A top 100 nm heavily doped n-type GaAs layer is capped for protection of underlying AlGaAs and to facilitate effective carrier injection as the n+ contact layer. The second grafted heterojunction (HJ2) is formed between the 80 nm p-type GaAsP and 250 nm n-type β-$Ga_2O_3$, with an ultrathin interfacial $Al_2O_3$ for double-side passivation and quantum tunneling, which are required for grafting.[26] Assuming a negligible potential drop at the $Al_2O_3$ layer, the band alignment of the n-p-n GaAs/AlGaAs/GaAsP/β-$Ga_2O_3$ double heterostructure under equilibrium is illustrated in Fig. 1(b). The bandgap values adopted for GaAs, AlGaAs, GaAsP and β-$Ga_2O_3$ are 1.42 eV, 1.80 eV, 1.84 eV, and 4.9 eV, respectively. Their respective electron affinity values are 4.07 eV, 3.74 eV, 3.80 eV, and 4.00 eV.

The growth of the β-$Ga_2O_3$ n++ sub-collector and n-type collector on an unintentionally doped (UID) β-$Ga_2O_3$ substrate was accomplished by plasma-assisted molecular beam epitaxy (PAMBE) using the Riber M7 system equipped with a Veeco $O_2$ bulb plasma source. The (010) oriented β-$Ga_2O_3$ UID substrate was cleaned of organics using a combination of acetone, methanol, and IPA before being loaded into the buffer of the system. The sample was baked at 400 °C for one hour to desorb any contaminants before being loaded into the main chamber for growth. The sample was heated to a substrate temperature of 725 °C before the oxygen plasma was struck at a plasma power of 250 W and an oxygen flow of 2.5 sccm as confirmed by a mass flow controller. A Ga flux of $1.5 \times 10^{-7}$ Torr as measured by nude filament gauge was used for the growth of the sub-collector. The growth rate at this condition is estimated to be ~4.5 nm/min from estimation of the separation between satellite peaks in an $\beta$-(Al$_x$Ga$_{1-x}$)$_2$O$_3$/$\beta$-Ga$_2$O$_3$ superlattice structure grown at the same condition. The growth of the 100 nm thick heavily doped n++ sub-collector was accomplished by introducing several delta doped layers by using a pulsed shuttering scheme for the Si effusion cell, which was at a temperature of 1050 °C. The shutter was opened for a period of 4 seconds followed by a period of 14 seconds closed, resulting in peak Si concentrations greater than $10^{20}$ cm$^{-3}$. The growth was stopped by shutting off the O$_2$ plasma and closing the Ga effusion cell shutter in order to move to the uniformly doped condition for the 250 nm thick n- collector. The growth was restarted with an increased plasma power of 300 W and reduced Ga flux of $1.1 \times 10^{-7}$ Torr, resulting in a Si concentration of $1 \times 10^{17}$ cm$^{-3}$ as confirmed by secondary-ion mass spectroscopy (SIMS). The background Si concentration at this condition is likely coming from impurities in the quartz plasma source as has been previously reported[1].

The fabrication of the AlGaAs/GaAsP/$\beta$-Ga$_2$O$_3$ double-heterojunction structure can be divided into two major stages, nanomembrane (NM) grafting and diode formation, as depicted in Fig. 2(a) and (b), respectively. During the initial grafting stage, the GaAs/AlGaAs/GaAsP epitaxial layers, as shown in Fig. 2(a i), was first patterned by 9×9 µm$^2$ mesh holes with a gap of 55 µm, using conventional photolithography. An inductively coupled plasma-reactive ion etching (ICP-RIE) process (Plasma Therm 770 ICP etcher, BCl$_3$: 10 sccm, and Ar: 5 sccm, pressure 15 mTorr, RIE power 60 W, and ICP power 500 W) was then employed to transfer the mesh hole pattern through the GaAs/AlGaAs/GaAsP layers, exposing the underlying sacrificial AlAs layer. The GaAs/AlGaAs/GaAsP NM was subsequently released by immersing the dry-etched epi in diluted hydrofluoric acid (HF:H$_2$O = 1:800) for 2 hours to chemically remove the sacrificial AlAs layer, as depicted in Fig. 1(a ii). After wet etching, the freestanding GaAs/AlGaAs/GaAsP NM, demonstrated in Fig. 1(a iii), was picked up from its source GaAs substrate using a polydimethylsiloxane (PDMS) stamp, preparing it for NM transfer printing, as illustrated in Fig. 1(a iv).

In parallel with the NM release process, contact metallization was performed on the n-/n+ $\beta$-Ga$_2$O$_3$ substrate prior to grafting, considering the potential thermal impact on the ultrathin interfacial oxide in subsequent fabrication processes. The metallization process started with dry etching of the top n- layer of the $\beta$-Ga$_2$O$_3$ substrate (Fig. 1(a I)) using the same ICP etcher (BCl$_3$: 18 sccm, Cl$_2$: 10 sccm, and Ar: 5 sccm, pressure 20 mTorr, RIE power 75 W, and ICP power 500 W). Upon exposure of the n+ $Ga_2O_3$ layer, cathode metal pads were deposited using a Ti/Au/Cu/Au metal stack (10/10/100/10 nm) through an electron beam evaporator (Angstrom Engineering Nexdep Physical Vapor Deposition Platform), followed by metal lift-off process and an ohmic rapid thermal annealing (RTA) at 600 °C for 10 seconds, as schematically shown in Fig. 2(a II). After completing the metallization step, the $\beta$-$Ga_2O_3$ substrate was cleaned by sonication in acetone, isopropyl alcohol (IPA), and deionized (DI) water for 10 minutes each, then dipped in buffered HF (HF:$H_2O$ = 1:10) for 1 minute. Directly after cleaning, the $\beta$-$Ga_2O_3$ substrate was put in a nitrogen bag and transferred into an ALD chamber, where 5 cycles of $Al_2O_3$ (~0.5 nm) were deposited at 250 °C. Following the ALD process, the released GaAs/AlGaAs/GaAsP NM was transfer printed to the n-region of the metallized $\beta$-$Ga_2O_3$ substrate with the assistance of a PDMS stamp. The grafting process was finalized with an annealing process at 350 °C for 5 minutes, allowing the formation of chemical bonding between the interfacial $Al_2O_3$ and the neighboring semiconductors, GaAsP and $\beta$-$Ga_2O_3$, resulting in the double-heterojunction structure, as schematically illustrated in Fig.2 (b i).

In the second stage of the double-heterojunction structure fabrication, another cathode metallization (Pd/Ge/Au/Cu/Au: 30/40/10/250/50 nm) was performed on the top n+ GaAs layer using the same e-beam evaporator, as shown in Fig.2 (b ii). By employing the metal on the n+ GaAs layer as an etching shadow mask, the GaAs/AlGaAs were dry etched at the exposed areas with the same ICP etcher and etching recipe, reaching the underlying p+ GaAsP layer, as displayed in Fig.2 (b iii). An anode metallization followed on the p+ GaAsP layer, using the metal stack of Ti/Pt/Au/Cu/Au: 15/50/20/300/100 nm, in the same e-beam evaporator. Afterwards, the GaAs/AlGaAs/GaAsP double heterojunction structures were isolated from each other by covering individual devices with photoresist (PR) serving as etching masks, and etched through the exposed membrane areas, as observed in Fig.2 (b v). After isolation, all the double-heterojunction devices were surface passivated with 80 cycles of $Al_2O_3$ (~8 nm) with the same ALD system. The final device structure is schematically demonstrated in Fig.2 (b vi).

An atomic force microscopy (AFM) image of the as-grown n-/n+ $\beta$-$Ga_2O_3$ substrate is presented in Fig. 3(a), displaying an ultralow surface roughness with a root mean square (RMS) value of only 0.67 nm within a scanning area of 5 × 5 $\mu m^2$. Fig. 3(b) presents a differential interference contrast (DIC) microscope image of the GaAs/AlGaAs/GaAsP NM transfer printed to $\beta$-$Ga_2O_3$ substrate upon completion of the chemical bonding process. As can be seen, the grafted NM is free of local defects such as wrinkles and air gaps, indicating a high-fidelity micro transfer process, as well as the well-preserved NM completeness. Fig. 3(c) illustrates the DIC microscopic images of the fabricated double-heterojunction structure with the interfingered electrodes of the HJ1 in Fig. 3(c i) representing the cathode on the AlGaAs layer and the anode on the GaAsP layer. Fig. 3(c ii) depicts a cathode metal pad formed on the exposed n+ $\beta$-$Ga_2O_3$ layer for ohmic contact. A three-dimensional optical scan of the device is displayed in Fig. 3(d), representing a three-layered device structure, consisting of AlGaAs, GaAsP, and $\beta$-$Ga_2O_3$ from top to bottom.

The I-V characterizations of the double-heterojunction devices were carried out using a Keithley 4200 Parameter Analyzer. Fig. 4 shows the measured I-V relationship for the n-p AlGaAs/GaAsP (marked in red) and the p-n GaAsP/$\beta$-$Ga_2O_3$ (marked in blue) heterojunctions. Both curves present comparable rectifying behaviors. The ON/OFF ratios at ±2 V for AlGaAs/GaAsP HJ1 and GaAsP/$\beta$-$Ga_2O_3$ HJ2 are $2.57 \times 10^3$ and $4.85 \times 10^2$, respectively, while the ideality factors are calculated to be 1.29 and 1.36, respectively. The low ON/OFF ratio of both heterojunctions can currently be attributed to the absence of appropriate thermal annealing for the AlGaAs cathode and GaAsP anode, leading to early influences of the series resistance and a limited forward current level. Despite the under-optimized fabrication process, the initial I-V measurements reveal that the grafted heterojunction exhibits equivalent performance to the epitaxial heterojunction.

In summary, this work demonstrates the potential of using semiconductor grafting technology to synthesize an epitaxy-like lattice-mismatched heterojunction of GaAsP/$\beta$-$Ga_2O_3$. This approach provides a viable solution to the challenge of deficient p-type doping in $\beta$-$Ga_2O_3$. Preliminary characterizations indicate that high-performance bipolar $\beta$-$Ga_2O_3$ devices for high-speed, high-power electronics, and optoelectronics can be anticipated in the future.

**Acknowledgments**

The work was supported by a CRG grant (2022-CRG11-5079.2) by the King Abdullah University of Science and Technology (KAUST). The work also received partial support from DARPA H2 program under grant: HR0011-21-9-0109.

## Figure Captions

**Fig. 1.** (a) Schematic illustration of the GaAs/AlGaAs/GaAsP/β-Ga$_2$O$_3$ n-p-n double-heterojunction (HJ) structure, with the epitaxial HJ1 comprising AlGaAs and GaAsP, and the grafted HJ2 formed between GaAsP and β-Ga$_2$O$_3$. (b) Band alignment of the GaAs/AlGaAs/GaAsP/β-Ga$_2$O$_3$ heterostructure at equilibrium.

**Fig. 2.** Process flow illustration of the GaAs/AlGaAs/GaAsP/β-Ga$_2$O$_3$ n-p-n double-heterojunction. (a) The grafting process, where GaAs/AlGaAs/GaAsP membrane is released from its source substrate and then transfer printed to the β-Ga$_2$O$_3$ substrate, which is pre-patterned with cathode pads and surface coated with Al$_2$O$_3$. (b) Device fabrication of the with a double mesa formation and triple metal deposition on respective layers. Each layer is represented by different colors, with their detailed materials indexed at the bottom of the figure.

**Fig. 3.** (a) An atomic force microscope (AFM) scanning of the as-grown β-Ga$_2$O$_3$ substrate. (b) Differential interference contrast (DIC) microscopic image of the GaAs/AlGaAs/GaAsP nanomembrane transferred on β-Ga$_2$O$_3$ substrate. (c) DIC microscopic images of fabricated devices, with i) interfingered cathode on GaAs/AlGaAs and anode on GaAsP, and ii) a cathode metal pad formed on n+ β-Ga$_2$O$_3$. (d) Three-dimensional optical profiling of the fabricated heterostructure.

**Fig. 4.** Measured I-V characteristics of the AlGaAs/GaAsP HJ1 and the GaAsP/β-Ga$_2$O$_3$ HJ2, presented on (a) the linear scale, and (b) the semi-logarithmic scale. The inset of (a) indicates the respective locations of HJ1 and HJ2, while the inset of (b) summarizes the extracted ON/OFF ratios and ideality factors from the measured curves.

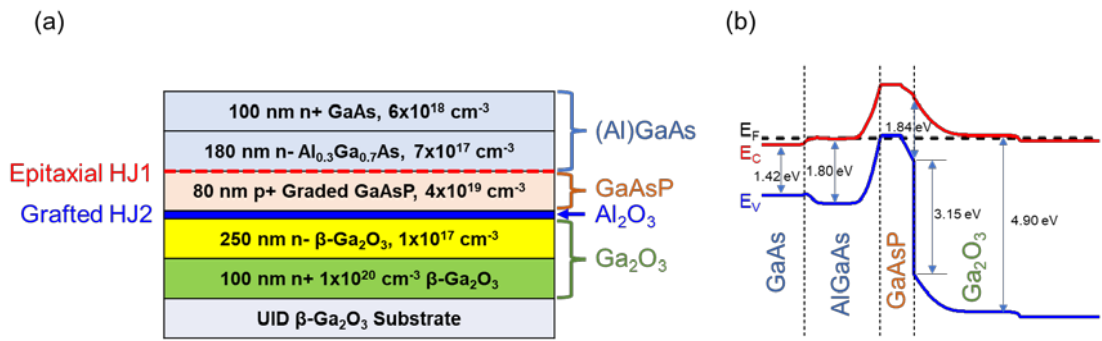

Fig.1.

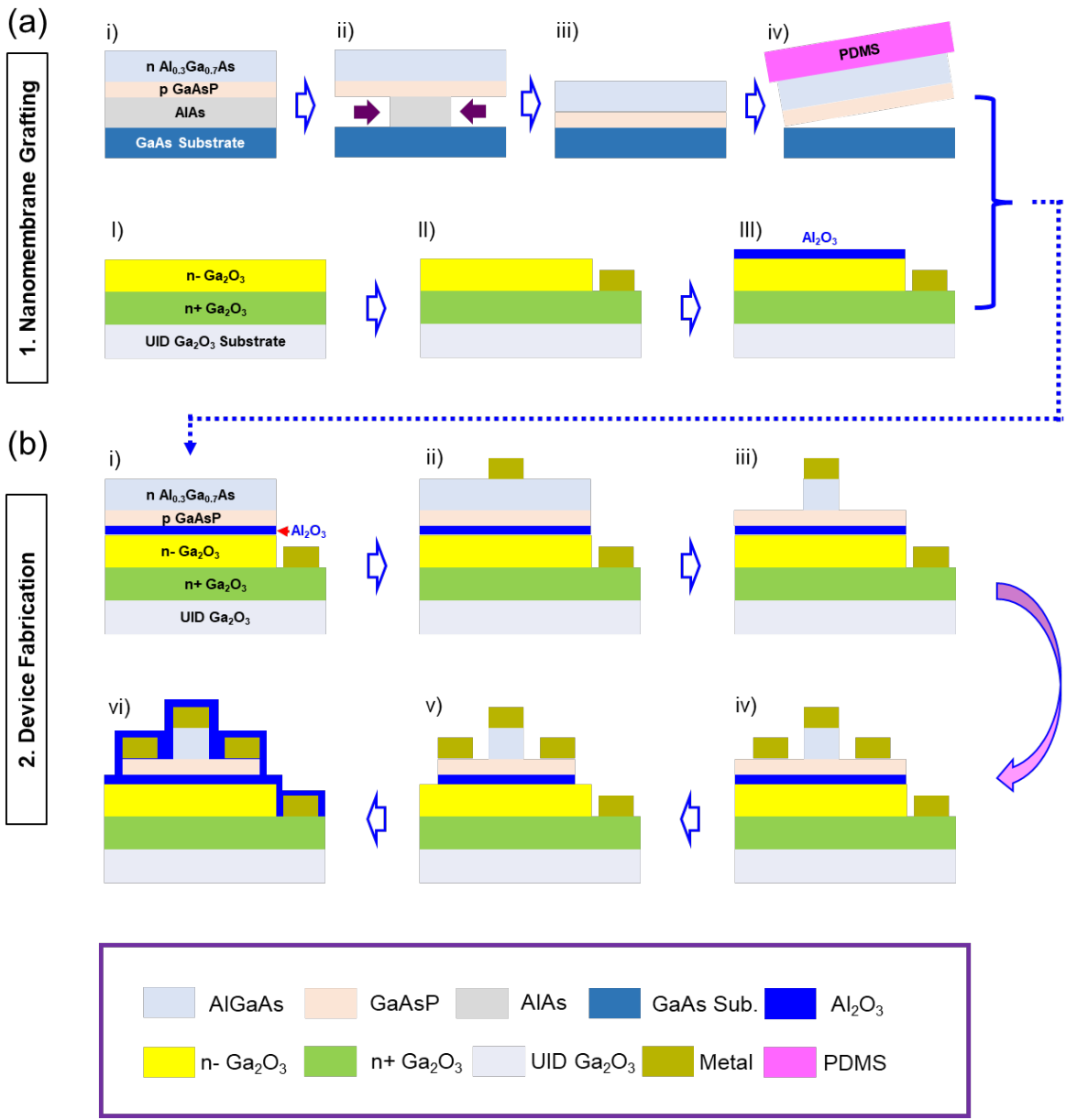

Fig. 2.

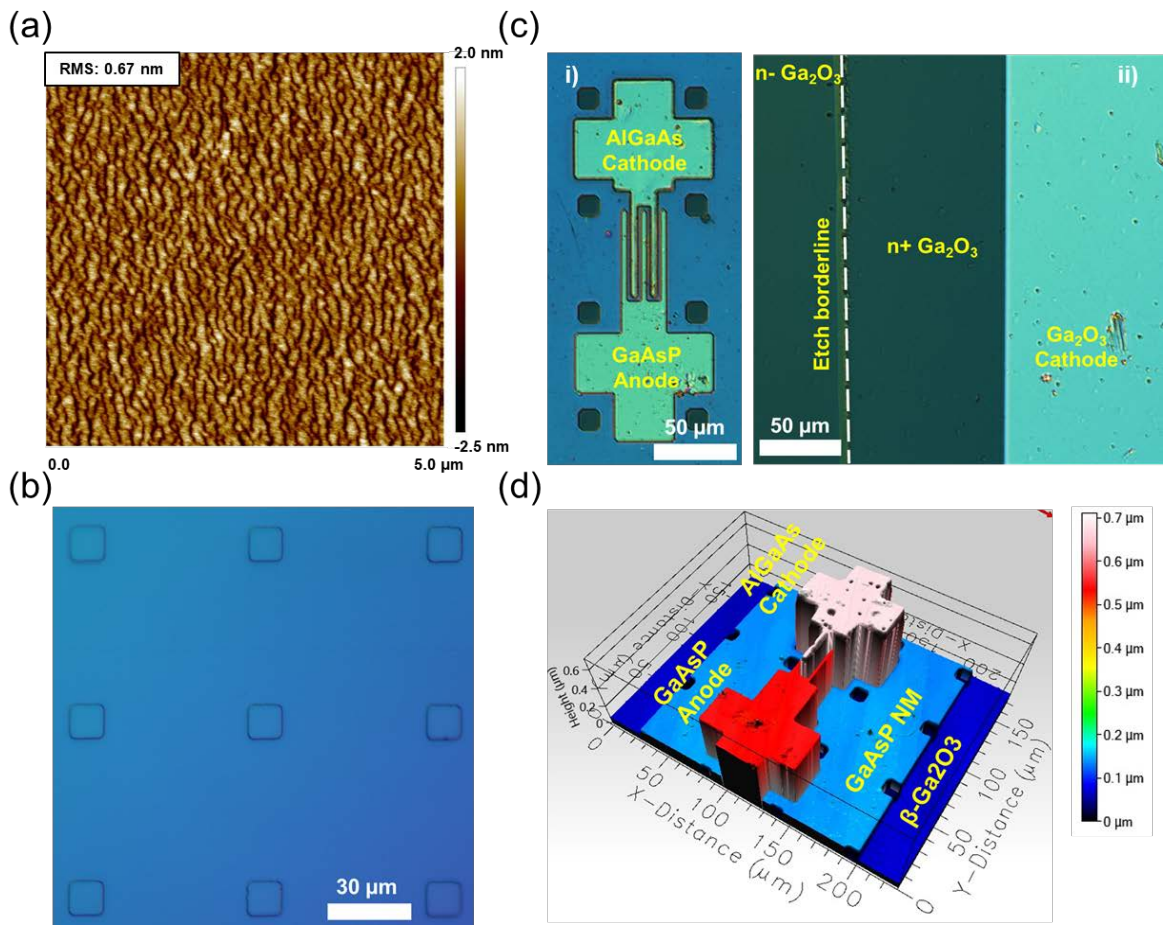

Fig. 3.

(a)

(b)

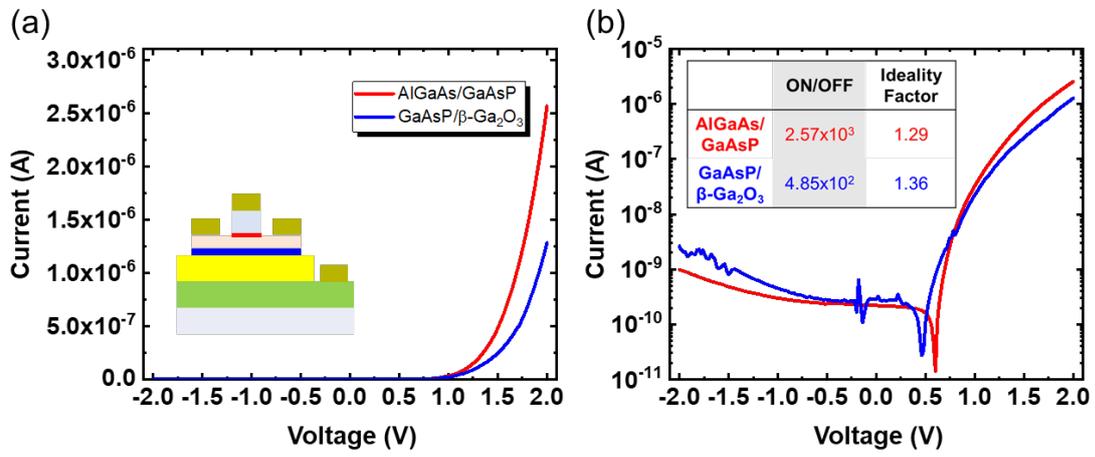

Fig. 4.